\documentclass[10pt]{iopart}
\usepackage{graphicx}
\usepackage[colorlinks=true, linkcolor=blue]{hyperref}
\usepackage{color, soul}
\usepackage{xcolor}

\usepackage[switch, columnwise]{lineno}
\expandafter\let\csname equation*\endcsname\relax
\expandafter\let\csname endequation*\endcsname\relax
\usepackage{amsmath}

\begin{document}

\title{Size and shape of skyrmions for variable Dzyaloshinskii-Moriya interaction and uniaxial anisotropy}

\author{Aroop Kumar Behera$^1$, Swapna Sindhu Mishra$^1$, Sougata Mallick$^1$, Braj Bhusan Singh$^1$, and Subhankar Bedanta$^1$}%
\address{$^1$Laboratory for Nanomagnetism and Magnetic Materials (LNMM), School of Physical Sciences, National Institute of Science Education and Research (NISER), HBNI, P.O.-Jatni, Khurda 752050, Odisha, India}%
\ead{sbedanta@niser.ac.in}
\vspace{10pt}
\begin{indented}
\item[18] April 2018
\end{indented}

\begin{abstract}
	We have performed micromagnetic simulations to study the formation of skyrmions in ferromagnetic elements with different shapes having perpendicular anisotropy. The strength of Dzyaloshinskii-Moriya interaction ($D$) and uniaxial anisotropy ($K$) are varied to elucidate the regime in which skyrmion formation can take place. It is found that for a certain combination of $D$ and $K$ skyrmion formation does not happen.   Further we also observed that for large $D$ and small $K$ values, finite size effect dominates which in turn hinders formation of typical N\'{e}el (spherical) skyrmions. However the resulting magnetic phase  is skyrmionic in nature and has different shape. 	  We also have found that the shape of the magnetic nano element has a significant role in determining the final magnetic state in addition to the competing $D$ and $K$ values. 
\end{abstract}



\maketitle

\ioptwocol

\section{Introduction:}
Skyrmions \cite{Skyrme_Skyrmion} are topological objects with a particular spin structure and are observed in ferromagnetic systems with prominent Dzyaloshinskii-Moriya interactions (DMI) \cite{Dzyaloshinskii,Moriya}. Skyrmions are classified as Bloch and N\'{e}el type\cite{Finocchio_Review}, predominantly observed in bulk materials and thin films, respectively. Skyrmions are also observed in wide range of materials like insulators \cite{Adams_Insulator}, doped semiconductors etc. \cite{Munzer_Doped}. It should be noted that thin films with perpendicular anisotropy and broken inversion symmetry may experience non-negligible interfacial DMI (iDMI) which leads to the formation of N\'{e}el skyrmions with radial chirality \cite{Heinz_Thinfilm}. 

Skymions in ultra-thin films are being extensively studied both experimentally and theoretically over the last decade. Recently it has been proposed that skyrmions in thin magnetic films have potential for future applications viz. in magnetic storage devices \cite{Kiselev_Storage} due to their stability and reduced dimensions. Skyrmion based devices are predicted to replace the conventional logic gates because driving skyrmions require lower current density \cite{Zhang_Logic}.  

The strength of iDMI ($D$) and anisotropy ($K$) determine the generation and size of skyrmions \cite{Zhang_Logic,Castro_Core,Chui_Skyrmion,Finnochio_Skyrmion,Jang_Mram,Novak_Skyrmion,Silva_Skyrmion}. The previous reports are mostly focused on a particular shape of the magnetic nanoelement e.g. nanodot or nanowire considering a set of values for $D$ and $K$. Recently a lot of attention has been focused on tuning $D$ and $K$ by choosing different combinations of ferromagnets (e.g. CoFeB, Co, Fe) and high spin orbit coupling materials (e.g. Pt, Au, W, Ir) for creating skyrmions \cite{Jiang_Skyrmion,Woo_Skyrmion,Grenz_Skyrmion,Hrabec_Skyrmion,Chaurasiya_DMI}.   Various experimental techniques such as Brillioun light scattering, spin polarized electron energy loss (SPEEL), etc. have been employed to evaluate the strength of DMI \cite{Chaurasiya_DMI,Zakeri_DMI}. However, it is experimentaly challenging to systematically tune the values of $D$ and $K$ to understand in which range of these two parameters skyrmions can be formed.  On the other hand, micromagnetic simulation provides unique advantage of choosing different combinations of $D$ and $K$ over a wide range \cite{Rohart_Nature,Nagaosa_Stable}. Recently, a lot of work regarding the observation of N\'{e}el skyrmions have been performed in multilayer structures with cobalt as the sandwich material\cite{Boulle_Co,Moreau_Co,Pollard_Co}. It has been observed that in Co/Pt or Co/Pd systems show strong spin-orbit coupling which is the prerequisite to obtain high iDMI. Additionally, the thickness of the Co layers can be tuned to modify the strength of $D$ and $K$. However inspite of several work related to this topic, a systematic study for the creation of skyrmions for various $D$ and $K$ in a wide range of values is necessary.\cite{Chui_Skyrmion,Finnochio_Skyrmion,Jang_Mram,Novak_Skyrmion} In this paper we have performed micromagnetic simulations on Co based systems to address this aspect using the object oriented micromagnetic framework (OOMMF) package \cite{Donahue_OOMMF}. Both the size and shape of the skyrmions strongly depend on the combination of $D$ and $K$ values and on the shapes of magnetic nanoelement. Further we have performed simulations by choosing material parameters of another ferromagnet i.e. Fe. We also compare our results to experimental observation of skyrmions. We believe our results will be helpful for experimentally designing magnetic multilayers for skyrmionic based devices.

Micromagnetic simulations solve the Landau-Lifshifz-Gilbert (LLG) equation:
\begin{equation}
\frac{d\textbf{m}}{dt}=-|\gamma|\textbf{m}\times\textbf{H}_{eff} + \alpha\left(\textbf{m} \times \frac {d\textbf{m}}{dt} \right)
\end{equation} to obtain the stable magnetization configuration in a ferromagnetic material \cite{Gilbert}. In equation (1), $\gamma$ is the gyromagnetic ratio and $\alpha$ is the Gilbert damping constant. $\textbf{H}_{eff}$ is the effective magnetic field of the system which is given by:
\begin{equation}
\textbf{H}_{eff}=-1/(\mu_oM_s)({\partial E_{total}}/{\partial \textbf{m}})
\end{equation}where, $\mu_o$ is the permeability in vacuum and \emph{$E_{total}$} is the magnetic energy of the system which consists of various energy terms such as exchange ($E_{ex}$), anisotropy ($E_{ani}$) and DMI ($E_{DMI}$). \cite{Liu_Chopping} Therefore, $E_{total}$ can be written as:

\begin{equation}
E_{total}=E_{demag} + E_{ex} + E_{ani} + E_{DMI}
\end{equation}
where, $E_{demag}$ is the demagnetization energy, $E_{ex} = -J_{i,j}{\bf{(\hat{{s}_i}\cdot\hat{{s}_j})}}$, $E_{ani}=K\bf{(\hat{{s}_i}\cdot\hat{z})^2}$, and $E_{DMI}=d_{ij}\cdot(\hat{s_i}\times\hat{s_j})$, respectively.   
Rohart \emph{et al.} have shown that  an extension to OOMMF is suitable to incorporate iDMI in thin-films for the study of skyrmions \cite{Rohart_Confine}. The contribution of DMI to the effective magnetic field is given by, 

\begin{equation}
\textbf{H}_{DMI}=\frac{2D}{\mu_{o}M_{s}}\left[(\nabla\cdot\textbf{m})\textbf{z}-\nabla m_z\right]
\end{equation}
where, \emph{D} is the DMI strength constant, \emph{$M_s$} is the saturation magnetization, $m_z$ is the magnetization component along z axis, $\textbf{m}$ is the magnetization and $\textbf{z}$ is the unit vector along z axis.

Another important parameter for a spin structure to be declared as a skyrmion is the skyrmion number or the topological number. The topological skyrmion number $N_{sk}$ is defined as the number of times $\textbf{m(r)}$ wraps around a unit sphere. For a skyrmion, the value of $N_{sk}$ is 1.
The skyrmion topological number in a thin film system is defined as\cite{Nagaosa_Stable}:
\begin{equation}
N_{sk}=\frac{1}{4\pi} \iint_{S} \textbf{m}\cdot \big({\frac{\partial \textbf{m}}{\partial x}}\times \frac{\partial \textbf{m}}{\partial y}\big)dxdy
\end{equation}
It should be noted that $N_{sk}$ is calculated over the whole magnetic element considered in the simulation.

\section{Experiment}

The 3D solver of the object oriented micromagnetic framework (OOMMF) \cite{Donahue_OOMMF} i.e. Oxsii was used for the simulations. As discussed earlier, an extension to OOMMF was used to introduce iDMI in the simulations \cite{Rohart_Confine}. We performed the micromagnetic simulations on a 0.6 nm thick Cobalt film with the area of 100 nm$\times$100 nm. We took the saturation magnetisation $M_{s}$ = 1.1$\times$$10^6$ A/m, and exchange stiffness A = 1.6$\times$$10^{-11}$ J/m for our system. The uniaxial anisotropy and the strength of the iDMI was varied for different simulations to check the optimum condition to achieve stable skyrmions. The cell size was kept at 0.5 nm$\times$0.5 nm$\times$0.6 nm. We have not used any periodic boundary conditions (PBC) in our simulations, because PBC would eliminate the contribution appearing from the edges of the nanoelements. The topological number of skyrmions was calculated by a home-built python code based on Equation 5. 

\section{Results and Discussion}

\begin{figure}[h!]
	\centering
	\includegraphics[width=1.0\linewidth]{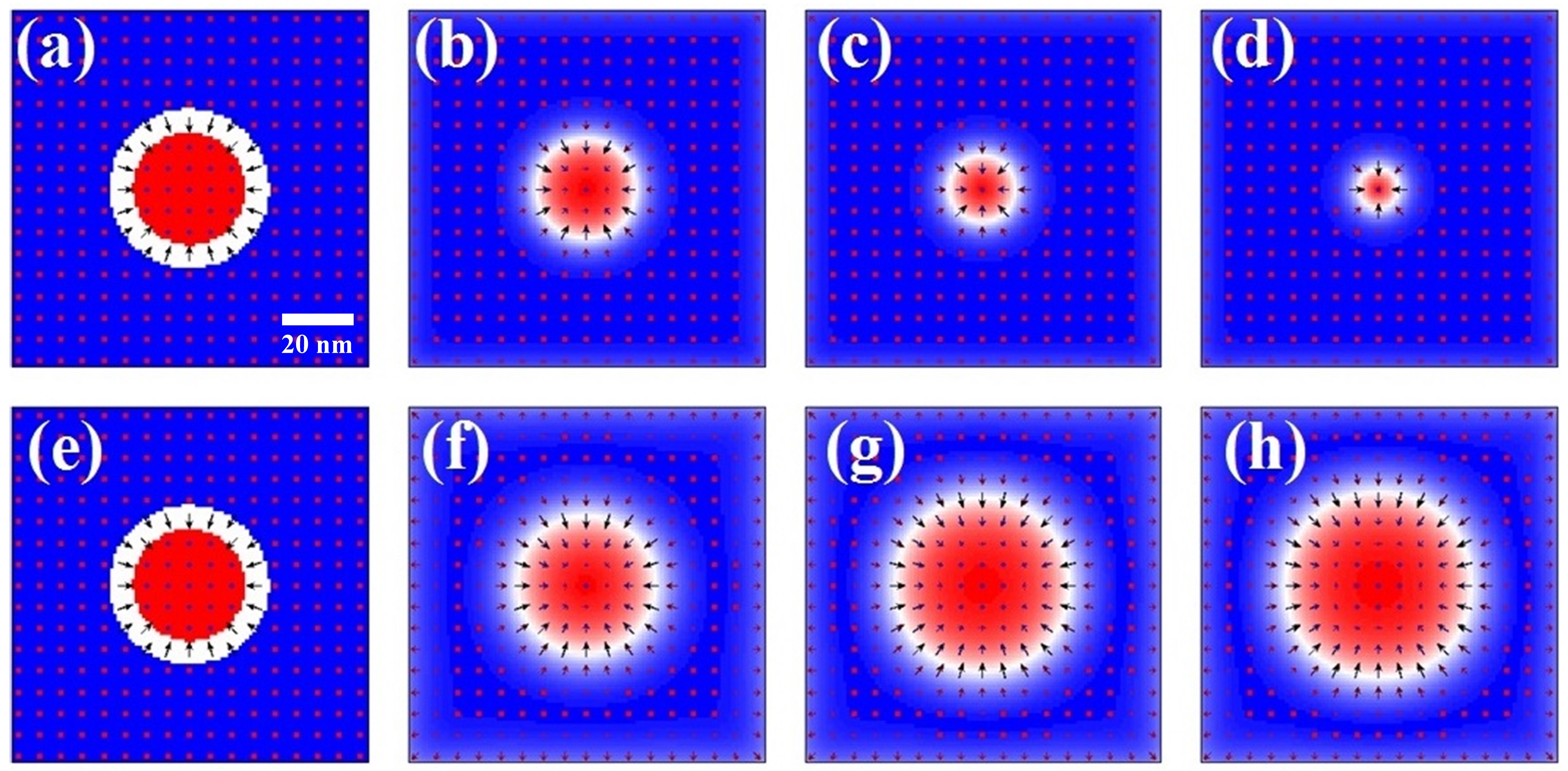}
	\caption{Creation and evolution of skyrmion from intitial bubble state for two representative simulations with $D$ = 4 mJ/$m^2$ and $K$ = 1$\times$$10^6$ J/$m^3$ (a)\textendash(d); and $D$ = 3.5 mJ/$m^2$ and $K$ = 0.4$\times$$10^6$ J/$m^3$ (e)\textendash(h), respectively. The simulation area of the ferromagnetic element is 100 nm$\times$100 nm. The initial bubble domain radius is same (22.5nm) for both the simulations.} 
	\label{fig:figure1}
\end{figure}

Figure \ref{fig:figure1} shows formation of typical N\'{e}el  skyrmions from a bubble domain state at the center of a ferromagnetic element (100 nm $\times$ 100 nm) without any external magnetic field. The inner and outer radii of the initial bubble domain are 16 and 22.5 nm, respectively. The bubble domain state has 2 distinct spin configurations with the red and blue area representing the +z and  $\textendash$z spin directions, respectively. The white area in the middle has spins in the xy plane which represents the width of the domain wall. It is expected that in the absence of any external magnetic field, the bubble would collapse due to its unstable nature. However, as the simulation begins, the influence of iDMI starts realigning the spins into a skyrmionic state where the change in spin direction from +z to $\textendash$z occurs continuously over a large area. The simulation ends when the spin configuration of the skyrmion becomes stable \cite{Rohart_Nature}. To ensure the stability of the final magnetic state, we have waited for sufficient time for completion of the simulation. We have used the criteria of $\frac{d\textbf{m}}{dt} <$ 0.01$^\circ/ns$. This states that when the change in angle of the tilt in the magnetization vector at any spatial point in the nanoelemnt is $<$ 0.01 degree/nano-second, the simulation will stop. The starting bubble domain and skyrmions look very similar to each other. However, one should note that, the spin rotation across a bubble domain and a skyrmion is different. For clarity the spin structure for bubble and a skyrmion is shown in Fig. S1 of the supplementary information. Further, the bubble is not a stable magnetic state and will evolve into either a skyrmionic, or a saturated state with time. Additionally, the skyrmions can be driven in a nanotrack with application of current pulses because of their topological stability whereas the bubbles will either collapse or saturate under it. Depending on the value of iDMI constant ($D$) and perpendicular uniaxial anisotropy ($K$), the stability and the sizes of the skyrmions get modified significantly. 
Figure \ref{fig:figure1}(a)\textendash(d) show the evolution of spins from the initial bubble state to a fully skyrmionic configuration  for $D$ = 4 mJ/$m^2$ and $K$ = 1$\times$$10^6$ J/$m^3$. In this particular simulation, the size of the skyrmion also starts to shrink to attain an energetically stable state. Similarly, Figure \ref{fig:figure1}(e)\textendash(h) shows the evolution of spins at different stages with $D$ = 3.5 mJ/$m^2$ and $K$ = 0.4$\times$$10^6$ J/$m^3$. It should be noted that in the later case the size of the skyrmion varies significantly in comparison to the former one (Figure \ref{fig:figure1}(d)) because of different $D$ and $K$. 

\begin{figure}[h!]
	\centering
	\includegraphics[width=1.0\linewidth]{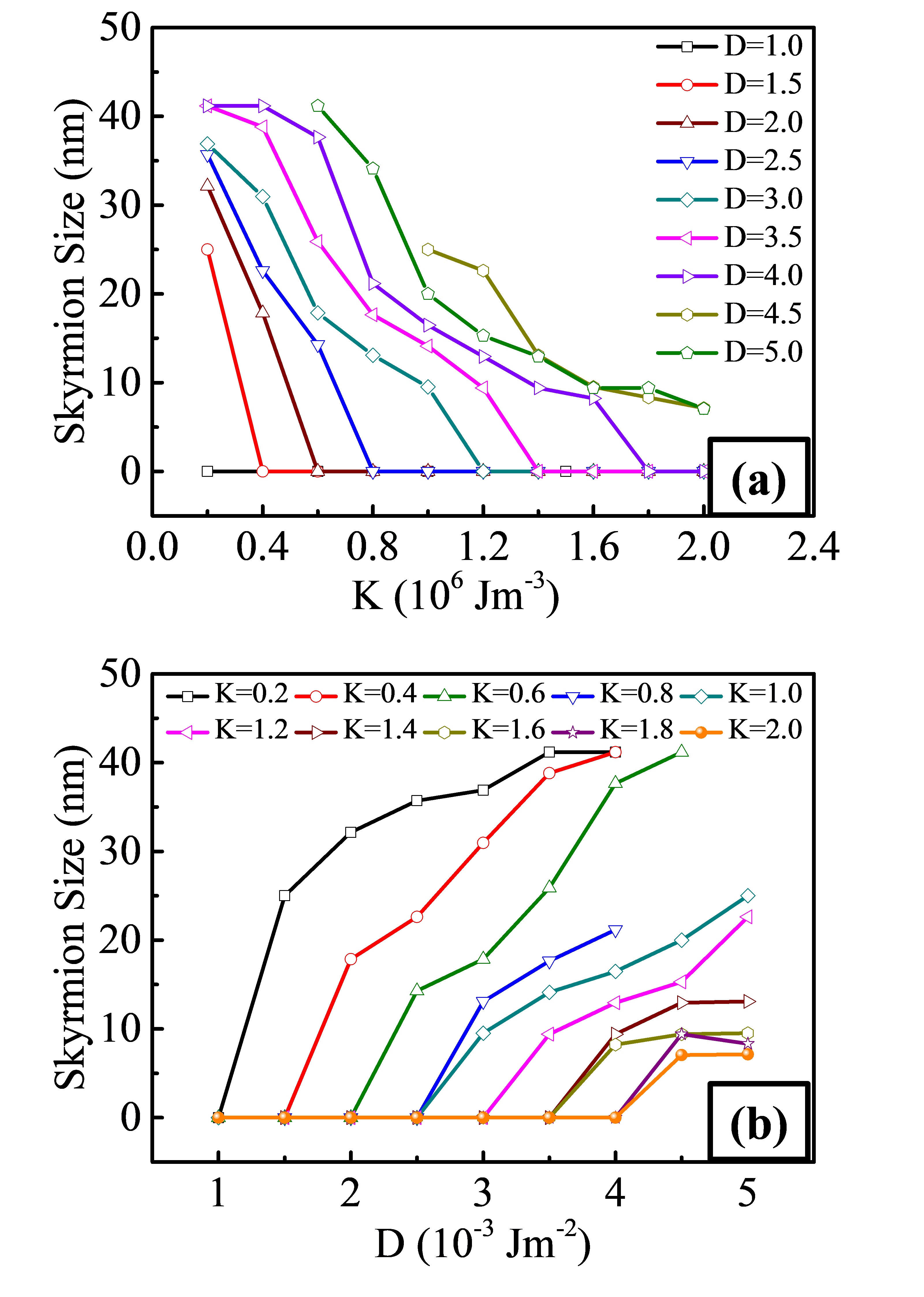}
	\caption{The size of skyrmions as a function of (a) $K$ (at different $D$), and (b) $D$ (at different $K$). The units of $D$ and $K$ in (a) and (b) are in mJ/$m^2$, and MJ/$m^3$, respectively.}
	\label{fig:figure2}
\end{figure}

Figure \ref{fig:figure2}(a) shows the variation of skyrmion radius with respect to $K$ for different values of $D$ represented by different colors. The skyrmion radius (referring to Fig. \ref{fig:figure1}) is defined as the distance between the end of $\textendash$z spin-orientation (blue region) to the centre of +z spin-orientation (red region) \cite{Rohart_Nature}. It shows that for a particular value of $D$ , the size of skyrmion decreases as the value of $K$ increases. Similarly, Figure \ref{fig:figure2}(b) shows the variation in skyrmion radius as a function of $D$ for different $K$ values. It is clear from Figure \ref{fig:figure2} that large values of $D$ and small values of $K$ is required for the formation of stable and large sized skyrmions.

\begin{figure*}
	\centering
	\includegraphics[width=1.0\linewidth]{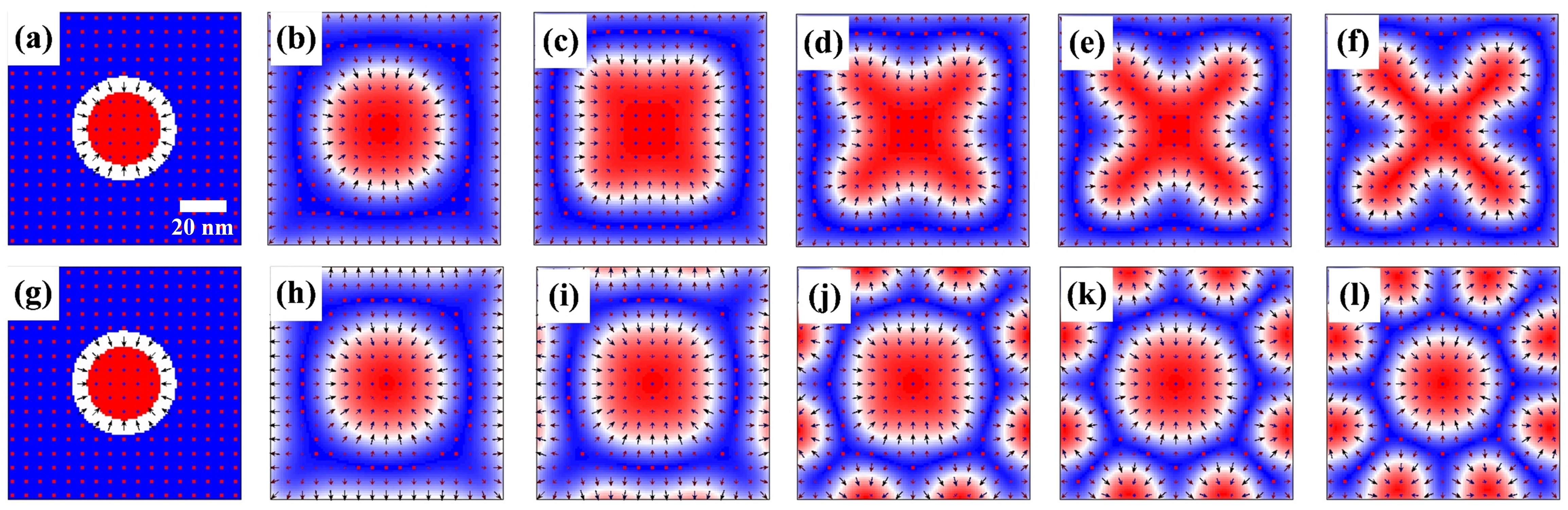}
	\caption {Simulations showing the development of skyrmionic state for $D$=5mJ/$m^2$ and $K$=0.4$\times$$10^6$ J/$m^3$ (a)-(f) and a helical state for $D$=5mJ/$m^2$ and $K$=0.2$\times$$10^6$ J/$m^3$ (g)-(l)}
	\label{fig:figure3}
\end{figure*}

In magnetic nanostructure it is well known that finite size effect plays a crucial role in determining the magnetic ground state. In device application it is desirable to study the existence of skyrmion formation in magnetic element with dimension of a few tens of nanometers. In this work we also explored the regime of $D$ and $K$ in which the finite size effect dominates.  Figure \ref{fig:figure3}(a)-(f) shows the development of an initial bubble state at $D$=5 mJ/$m^2$ and $K$=0.4$\times$$10^6$ J/$m^3$. As mentioned earlier, the topological number($N_{sk}$) = 1 of a stable magnetic state corresponds to a skyrmionic state. $N_{sk}$ has been derived using a python code based on eqn. 5. It should be noted that $N_{sk}$  $\approx$ 1, for this particular state, indicates that the stable state is not completely skyrmionic and has unusual shape due to the edge effect. The type of the magnetic state having the shape shown in figure \ref{fig:figure3}(f) is called as incomplete skyrmion \cite{Pepper_Skyrmion,Beg_Skyrmion}. Further, figure \ref{fig:figure3}(g)-(l) shows similar development of magnetic states at $D$=5 mJ/$m^2$ and $K$=0.2$\times$$10^6$ J/$m^3$. For the magnetic state shown in figure \ref{fig:figure3}(l) $N_{sk}\approx5$. Such magnetic configurations with higher $N_{sk}$ values are reported as helical state in literature \cite{Beg_Skyrmion}. However, one should note that, if the $N_{sk}$ is calculated individually for the closed topological object at the center, it would lead to $N_{sk}=1$ (for Fig. \ref{fig:figure3}(l)). Therefore, the magnetic states observed under the edge effect can be divided into two parts: incomplete skyrmions and helical states. These two type of stable magnetic structures strongly rely on the interplay of the values of $D$ and $K$ as well as the finite size effect of the magnetic nanoelement. 

\begin{figure}[h!]
	\centering
	\includegraphics[width=0.8\linewidth]{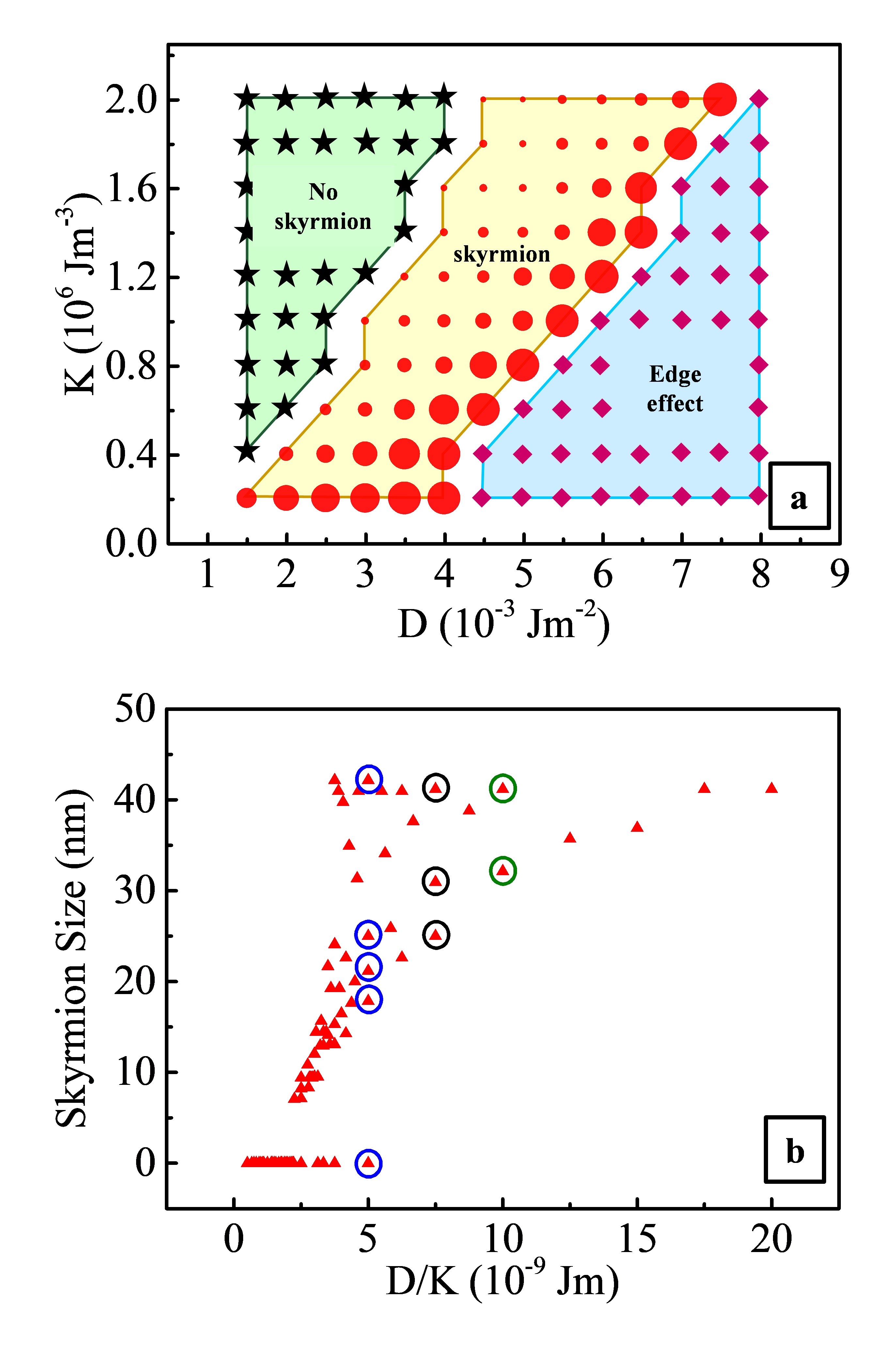}
	\caption {(a) Phase plot showing the possibility of existence of {\bf skyrmions} along with other magnetic states (magnetic saturation state achieved with {\bf no skyrmion}, helical states dominated by {\bf edge effect}) depending on the values of $D$ and $K$. The red circular dots represent the relative size of the skyrmions normalized to the maximum size (i.e., 41.2 nm) which we obtained. The star data points represent the magnetic state with no skyrmions. The data points represented by diamond (pink) depict the magnetic states dominated by edge effect. (b) Variation of the radius of the skyrmions with respect to the $D$/$K$ ratio which is analyzed from fig 4a. The blue, black, and green circles represent data points with $D$/$K$ ratio of 5, 7.5, 10 nJm, respectively, with different $D$ and $K$ values (refer to main text for detailed explanation).}
	\label{fig:figure4}
\end{figure}

Figure \ref{fig:figure4}(a) shows a phase map of different magnetization states (saturation, skyrmion, domains dominated by edge effects) evolved from a bubble domain under the influence of the strength of iDMI and anisotropy varied over a wide range. It is observed that the bubble domain evolves into a saturated state when the value of $D$  is very low. The green colored area marked in Figure \ref{fig:figure4}(a) shows the data points (black star) with different $D$ and $K$ values, where no skyrmion is generated and with time the bubble evolves into saturation. Simulations reveal that, 1.5 $\leq$ $D$ $\leq$ 7.5 mJ/$m^2$, and $K$ $\geq$ 0.2 MJ/$m^3$ are the pre-requisite for the formation of skyrmions. The magnitude of $D$ and $K$ also govern the radius of the skyrmions as shown by the red circular dots with different sizes on a relative scale. The smallest and largets radii of the skyrmions are 7.1 nm and 41.2 nm, respectively. The size of the skyrmions from our work matches well to the experimental findings of previously reported literature \cite{Fert_Track,Romming_Skyrmions,Yu_Skyrmion,Seki_Skyrmion}. We observed that when the value of $D$ is $\geq$ 4.5 mJ/$m^2$, for lower value of $K$, the bubbles evolve into helical domain states \cite{Rohart_Confine} and domain states dominated by the edge effect \cite{Muller_Skyrmion,Ran_Edge}. The area marked with cyan color shows the data points (pink diamonds) with different values of $D$ and $K$ where the bubbles evlove into skyrmionic and helical states with edge effects. In order to understand the dependence of skyrmion size on the values of $D$ and $K$, we plotted the radii of skyrmions as a function of $D$/$K$ in Figure \ref{fig:figure4}(b). The blue circled data points in Figure \ref{fig:figure4}(b) show the radii of the skyrmions for the value of $D$/$K$=5.0 nJm (for different values of $D$ and $K$). The values of the radii obtained at $D$/$K$=5.0 nJm are 0 nm (no skyrmion)($D$=1.0 mJ/$m^2$ , $K$=0.2 MJ/$m^3$), 17.9 nm ($D$=2.0 mJ/$m^2$ , $K$=0.4 MJ/$m^3$), 21.2 nm ($D$=4.0 mJ/$m^2$ , $K$=0.8 MJ/$m^3$), 25.0 nm ($D$=5.0 mJ/$m^2$ , $K$=1.0 MJ/$m^3$), and 42.2 nm ($D$=6.0 mJ/$m^2$ , $K$=1.2 MJ/$m^3$). Similarly, the black and green circled data points in Fig. \ref{fig:figure4}(b) represent the skyrmions sizes obtained at $D$/$K$=7.5 nJm and $D$/$K$=10.0 nJm, respectively. The radius obtained for $D$/$K$=7.5 nJm are 25.0 nm ($D$=1.5 mJ/$m^2$ , $K$=0.2 MJ/$m^3$), 30.9 nm ($D$=3.0 mJ/$m^2$ , $K$=0.4 MJ/$m^3$), and 41.2 nm ($D$=4.5 mJ/$m^2$ , $K$=0.6 MJ/$m^3$). Further, the radius of the skyrmions for $D$/$K$=10.0 nJm are 32.1 nm ($D$=2.0 mJ/$m^2$ , $K$=0.2 MJ/$m^3$), and 41.2 nm ($D$=4.0 mJ/$m^2$ , $K$=0.4 MJ/$m^3$). From this comparison, we note that for the same $D/K$ ratio the radius of the skyrmions can be different. Similar but few simulations were carried out by taking the material parameters of Iron. A similar behaviour towards the relaxation of bubble domains to form skyrmions was also observed (see supplementary Fig.S2).

\begin{figure}[h!]
	\centering
	\includegraphics[width=1.0\linewidth]{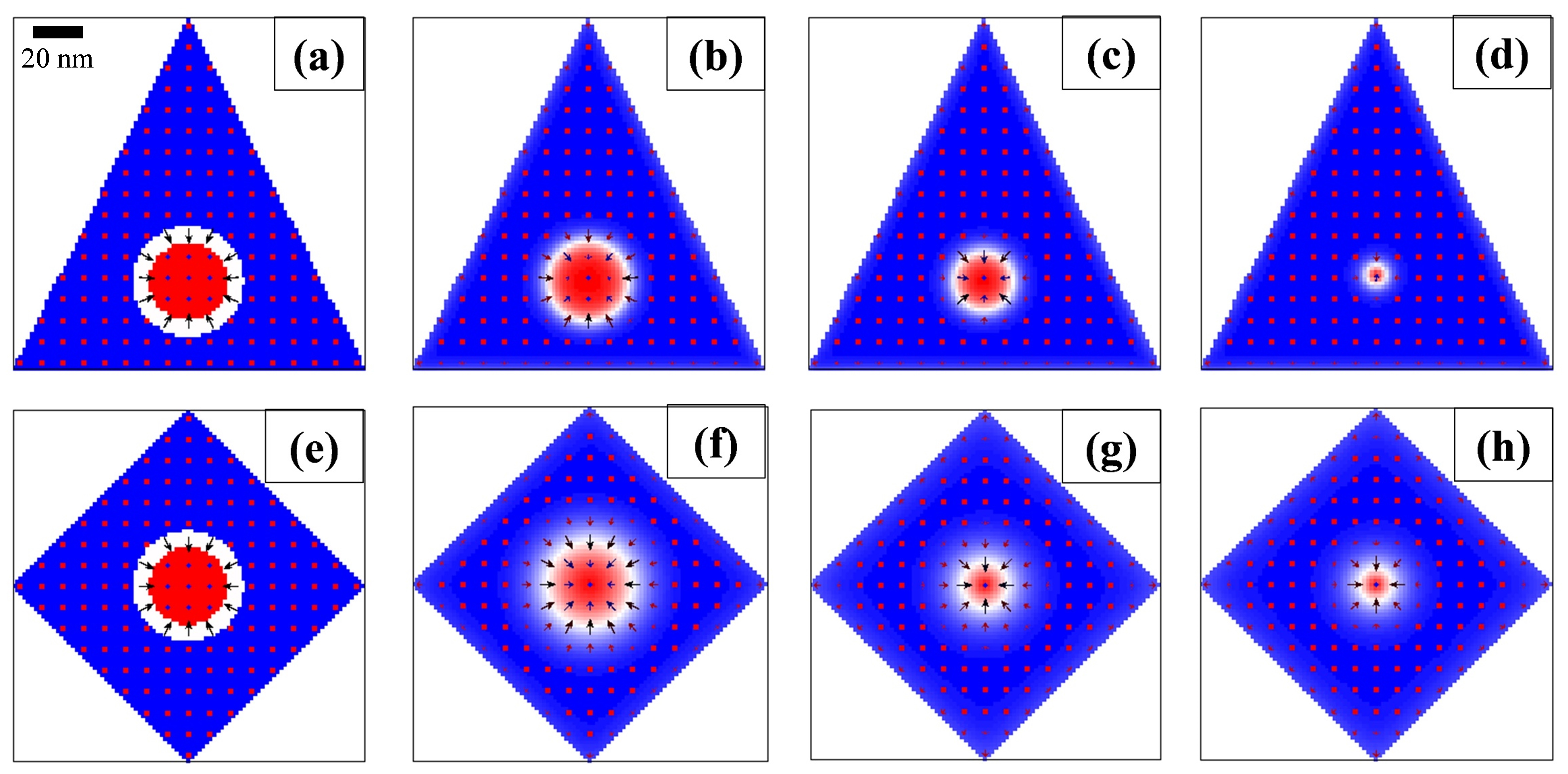}
	\caption {(a)-(d) Creation and evolution of skyrmion from initial bubble state for two representative simulations with $D$=4.5 mJ/$m^{2}$ and $K$=1.4 MJ/$m^3$. (e)\textendash (h) showing the stepwise development of skyrmion from bubble domain state in a diamond shaped sample for material parameters $D$=1.5 mJ/$m^2$ and $K$=0.2 MJ/$m^3$.}
	\label{fig:figure5}
\end{figure}

Further, to study the role of shape of the magnetic nanoelement in confining skyrmions, we simulated various types of nanoelements such as triangular and diamond. In all the simulations, the net sample area was kept equal to that of the square nanoelements shown in Fig.\ref{fig:figure1}-\ref{fig:figure3}. Fig. \ref{fig:figure5}(a)\textendash (d) show the development of a bubble domain in a triangular sample with $D$=4.5 mJ/$m^{2}$ and $K$=1.4 MJ/$m^3$. Fig.\ref{fig:figure5}(e)\textendash (h) show similar development of skyrmion from a bubble domain confined in a diamond shaped sample for $D$=1.5 mJ/$m^2$ and $K$=0.2 MJ/$m^3$. Nevertheless, there is no significant difference between a diamond and a square shaped magnetic element, apart from the fact that it is 45$^\circ$ rotated. However, in a periodic lattice of such nanoelements, the shape anisotropy arising from the two different elements will be different. Hence, we conclude that by choosing proper combination of $D$ and $K$, skyrmions can be stabilized in different shaped nanostructures.

\begin{figure}[h!]
	\centering
	\includegraphics[width=0.75\linewidth]{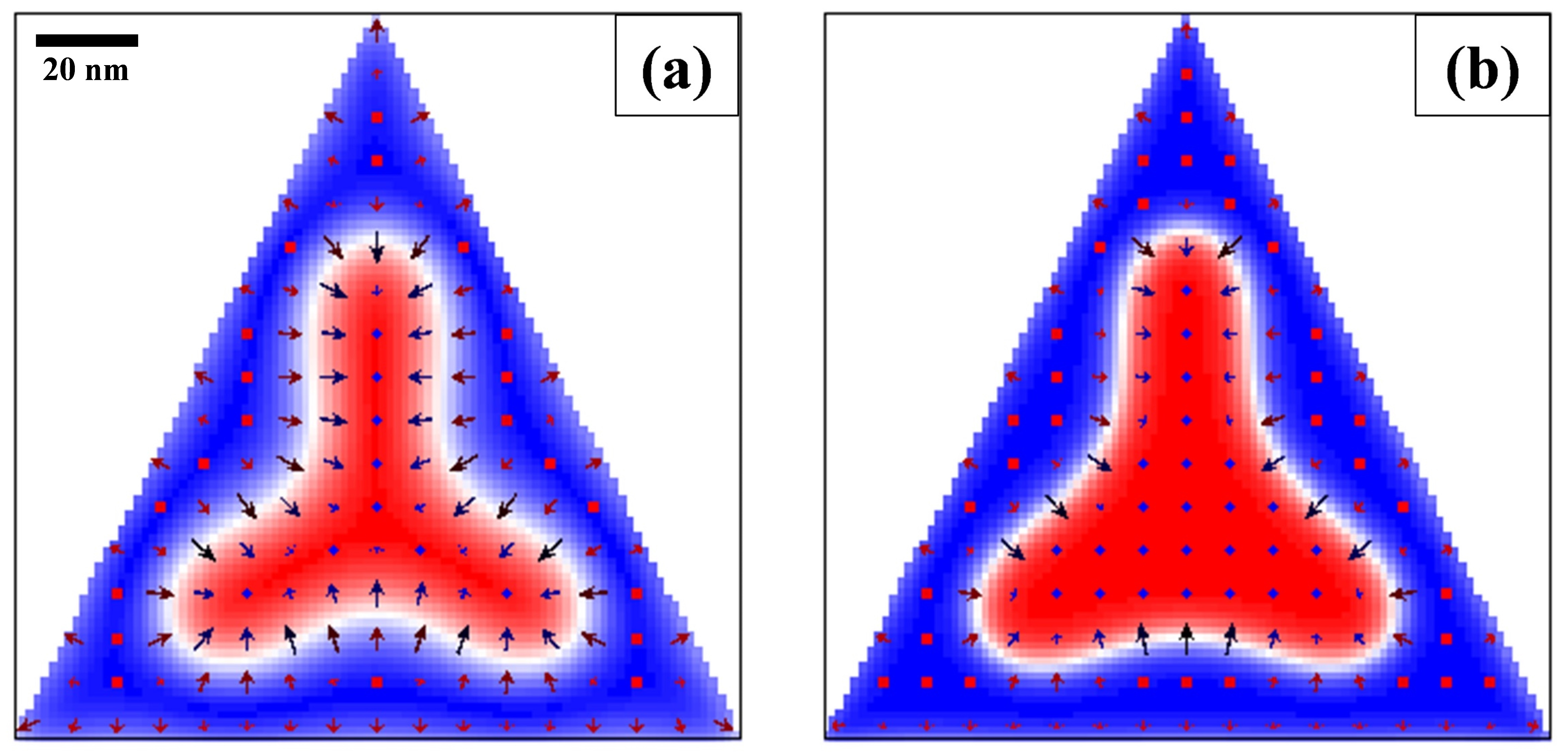}
	\caption {The final stable states in a triangular nanoelement  are shown for $D$=4.5 mJ/$m^2$ and $K$=0.4 MJ/$m^3$ (a) and $D$=7.5 mJ/$m^2$ and $K$=2.0 MJ/$m^3$ (b). These states are called incomplete skyrmion  because $N_{sk}$ $\simeq$ 1.}
	\label{fig:figure6}
\end{figure}

Fig.\ref{fig:figure6}(a) and \ref{fig:figure6}(b) show two special domain strutures obtained in the triangular sample for $D$=4.5 mJ/$m^2$, $K$=0.4 MJ/$m^3$; and $D$=7.5 mJ/$m^2$, $K$=2.0 MJ/$m^3$, respectively. It should be noted that, in spite of the unconventional shape, calculation of topological number in such domain strutures yielded the value $N_{sk}$ $\approx$1. Hence these magnetic states can be called as a incomplete skyrmion.

\begin{figure}[h!]
	\centering
	\includegraphics[width=0.75\linewidth]{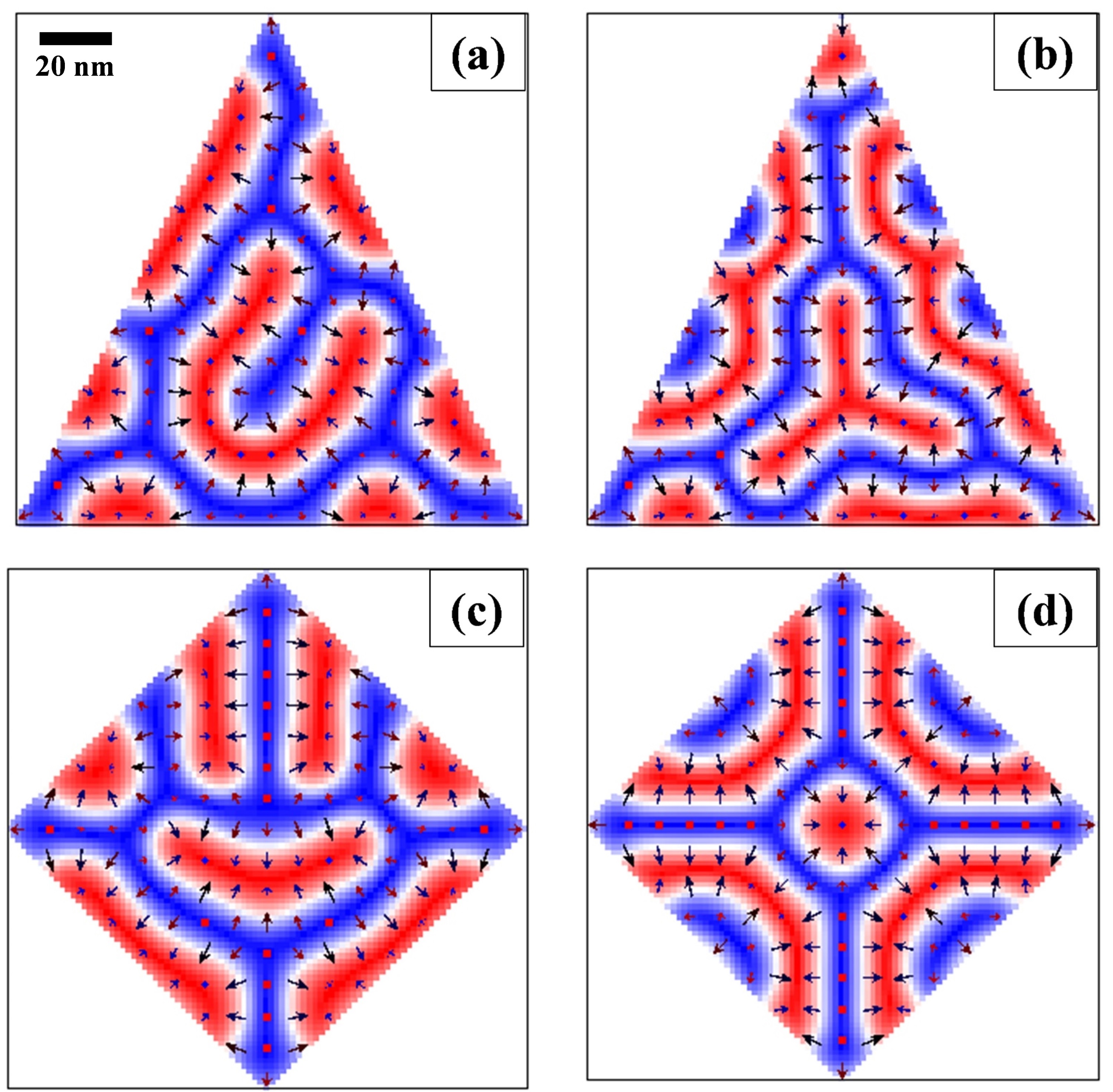}
	\caption {Formation of various helical states from bubble domains trapped in various samples of different shapes, sizes and magnetic properties. Material parameters are (a) $D$=7.0 mJ/$m^2$ and $K$=0.6 MJ/$m^3$, (b)  $D$=8.0 mJ/$m^2$ and $K$=0.2 MJ/$m^3$, (c) $D$=7.0 mJ/$m^2$ and $K$=0.6 MJ/$m^3$ and (d) $D$=8.0 mJ/$m^2$ and $K$=0.2 MJ/$m^3$.}
	\label{fig:figure7}
\end{figure}

We also observe formation of helical states (similar to \ref{fig:figure3}(g)\textendash (l)) in triangular and diamond shaped samples. Fig.\ref{fig:figure7} show such helical state in the triangular sample for (a) $D$=7.0 mJ/$m^2$, $K$=0.6 MJ/$m^3$, and (b) for $D$=8.0 mJ/$m^2$, $K$=0.2 MJ/$m^3$. Similarly, fig.\ref{fig:figure7} (c) and (d) show the helical states observed in the diamond sample for $D$=7.0 mJ/$m^2$, $K$=0.6 MJ/$m^3$; and $D$=7.0 mJ/$m^2$, $K$=0.6 MJ/$m^3$, respectively.

\section{Conclusions}

The stability of single skyrmion in the presence of iDMI has been discussed with the help of micromagnetic simulations. Extensive simulations show that the size and shape of N\'{e}el skyrmion strongly depend on the iDMI strength as well as on the perpendicular anisotropy of the material. In the experiments the $K$ and $D$ values may be chosen systematically by varying the thickness of the constituent layers to obain skyrmions with very low dimension from applications point of view. We also note that achieving very high value of $D$ is not feasible from experimental point of view. This aspect can be further taken care of by choosing low perpendicular anisotropy of the magnetic heterostructure. Further it is found that for a set of $D$ and $K$ values the finite size of the magnetic nanoelement plays a crucial role. The latter may lead to helical domain states or even to a skyrmionic stable state which has a shape very different as compared to the conventional (i.e. circular) N\'{e}el skyrmions. Further studies are required to elucidate the effect of the finite size and shape of the magnetic nanostructure on the formation of such unusual skyrmionic states. The nature of the magnetic state for a certain $D$ and $K$ combination is similar in different shapes of magnetic nanoelements studied in this work. This further confirms that the combination of $D$ and $K$ values plays the driving role in determining the magnetic configuration in such systems. The results presented in this work may be considered as a phase diagram for designing experimental samples (thin film heterostructures) for creating skyrmions which can be of significant importance in future spintronic applications.

\section*{Acknowledgements}

We thank Department of Atomic Energy (DAE) and Indo-French collaborative project supported by CEFIPRA, Govt. of India for providing the funding to carry out the research work. We acknowledge the technical help of Mr. Siddharth Mansingh for developing the python code to calculate the topological number of skyrmions. We also thank Prof. Ajaya Nayak for proof reading the manuscript.


\section*{References}
\begin{thebibliography}{99}
	
\bibitem{Skyrme_Skyrmion} Skyrme T H R 1962 \emph{Nucl. Phys.} {\bf 31} 556-569

\bibitem{Dzyaloshinskii} Dzyaloshinskii I 1958 \emph{J. Phys. Chem. Solids } {\bf 4} 241

\bibitem{Moriya} Moriya T 1960 \emph{Phys. Rev. } {\bf 120} 91 

\bibitem{Finocchio_Review} Finocchio G, B$\ddot{u}$ttner F, Tomasello R, Carpenntieri M and  Kl$\ddot{a}$ui M 2016 \emph{J. Phys. D: Appl. Phys.} {\bf 49} 423001 

\bibitem{Adams_Insulator} Adams T, Chacon A, Wagner M, Bauer A, Brandl G, Pedersen B, Berger H, Lemmens P and Pfleiderer C 2012 \emph{Phys. Rev. Lett.} {\bf 108}  237204

\bibitem{Munzer_Doped} M$\ddot{u}$nzer W \emph{et al} 2010 \emph{Phys. Rev. B } {\bf 81} 041203 

\bibitem{Heinz_Thinfilm} Heinze S, von Bergmann K, Menzel M, Brede J,  Kubetzka A, Wiesendanger R,  Bihlmayer G and Blugel S 2011 \emph{Nat. Phys.} {\bf 7} 713

\bibitem{Kiselev_Storage} Kiselev N S, Bogdanov A N, Sch$\ddot{a}$fer R and  R$\ddot{o}$bler U K 2011 \emph{J. Phys. D: Appl. Phys.} {\bf 44 } 392001

\bibitem{Zhang_Logic} Zhang X, Ezawa M and Zhou Y 2015 \emph{Sci. Rep.} {\bf 5} 9400 

\bibitem{Castro_Core} Castro M A and Allende S 2016 \emph{J. Magn. Magn. Mater.}  {\bf 417} 344

\bibitem{Chui_Skyrmion} Chui C P, Ma F and Zhou Y 2015 \emph{AIP Adv.} {\bf 5} 047141

\bibitem{Finnochio_Skyrmion} Finocchio G, Ricci M, Tomasello R, Giordano A, Lanuzza M, Puliafito V, Burrascano P, Azzerboni B and Carpentieri M 2016 \emph{Appl. Phys. Lett.}  {\bf 107} 262401 

\bibitem{Jang_Mram} Jang P H, Song K, Lee S J, Lee S W, Lee K J 2015 \emph{Appl. Phys. Lett.} {\bf 107} 202401

\bibitem{Novak_Skyrmion} Novak R L, Garcia F, Novais E R P, Sinnecker J P and Guimarães A P  2018 \emph{J. Magn. Magn. Mater.} {\bf 451} 749 

\bibitem{Silva_Skyrmion} Vidal-Silva N, Riveros A and Escrig J 2017 \emph{J. Magn. Magn. Mater.} {\bf 443} 116

\bibitem{Jiang_Skyrmion}Jiang W, Upadhyaya P, Zhang W, Yu G, Jungfleisch M B, Fradin F Y, Pearson J E, Tserkovnyak Y, Wang K L,Heinonen O, te Velthuis S G E and Hoffmann A 2015 \emph{Science} {\bf 349} 283

\bibitem{Woo_Skyrmion} Woo S, Litzius K, Kr$\ddot{u}$ger B, Im M Y, Caretta L, Richter K, Mann M, Krone A, Reeve R M, Weigand M, Agarwal P, Lemesh I, Mawass M A, Fischer P, Kl$\ddot{a}$ui M and Beach S D 2016 \emph{Nat. Mater.} {\bf 15} 501

\bibitem{Grenz_Skyrmion} Grenz J, K$\ddot{o}$hler A, Schwarz A and Wiesendanger R 2017 \emph{Phys. Rev. Lett.} {\bf 119} 047205

\bibitem{Hrabec_Skyrmion} Hrabec A, Sampaio J, Balmeguenai M, Gross I, Weil R, Cherif S M, Stashkevich A, Jacques V, Thiaville A and Rohart S 2017 \emph{Nat. Comm.} {\bf 8} 15765

\bibitem{Chaurasiya_DMI} Chaurasiya A K, Banerjee C, Pan S, Sahoo S, Choudhury S, Sinha J and Barman A 2016 \emph{Sci. Rep.} {\bf 6} 32592  

\bibitem{Zakeri_DMI} Zakeri K, Zhang Y, Prokop J, Chuang T H, Sakr N, Tang W X and Kirschner J 2010 \emph{Phys. Rev. Lett.} {\bf 104} 137203 

\bibitem{Rohart_Nature} Sampaio J, Cros V, Rohart S, Thiaville A and Fert A 2013 \emph{ Nat. Nanotechnol.} {\bf 8} 839

\bibitem{Nagaosa_Stable} Nagaosa N and Yoshinori T 2013  \emph{Nat. Nanotechnol.} {\bf 8} 899

\bibitem{Boulle_Co} Boulle O, Vogel J, Yang H, Pizzini S, Chaves DS, Locatelli A, Mentes T O, Sala A, Buda-Prejbeanu L D, Klein O, Belmeguenai M, Roussigne Y, Stashkevich A, Cherif S M, Aballe L, Foerster M, Chshiev M, Auffret S, Miron I M and Gaudin G 2016 \emph{Nat. Nanotech.}{\bf 11} 449

\bibitem{Moreau_Co} Moreau-Luchaire, Moutafis C, Reyren J, Sampaio J, Vaz C A F, Van Horne N, Bouzehousane K, Garcia K, Deranlot C, Warnicke P, Wohl$\ddot{u}$ter P, George J M, Weigand M, Raabe J, Cros V and Fert A 2013 \emph{Nat. Nanotech.} {\bf 11} 444

\bibitem{Pollard_Co} Pollard D S, Garlow J A, Yu J, Wang Z, Zhy Y and Yang H 2017 \emph{Nat. Commn.} {\bf 8} 14761

\bibitem{Donahue_OOMMF} Donahue M J and Porter D G 1999 \emph{Interagency Report \textbf{NISTIR}} {\bf 6376} 

\bibitem{Gilbert} Gilbert A and Thomas L 2004  \emph{IEEE Trans. Magn.} {\bf 40} 3443

\bibitem{Liu_Chopping} Liu Y, Lei N, Zhao W, Liu W, Ruotolo A, Braun H B and Zhou Y 2017 \emph{Appl. Phys. Lett.} {\bf 111} 022406

\bibitem{Rohart_Confine} Rohart S and Thiaville A 2013  \emph{Phys. Rev. B} {\bf 88 } 184422

\bibitem{Pepper_Skyrmion} Pepper R A, Beg M, Cortes-Ortuno D, Kluyver T, Bisotti M A, Carey R, Albert M, Wang W, Hovorka O and Fangohr H 2018 \emph{J. Appl. Phys} {\bf 123} 093903

\bibitem{Beg_Skyrmion} Beg M, Carey R, Wang W, Cortes-Ortuno D, Vousden M, Bisotti M A, Albert M,Chernyshenko D, Hovorka O, Stamps R L and Fangohr H 2015 \emph{Sci. Rep.} {\bf 5} 17137

\bibitem{Fert_Track} Fert A, Cros V and Sampaio 2013 J \emph{Nat. Nanotech.} {\bf 8} 152

\bibitem{Romming_Skyrmions} Romming N, Hanneken C, Menzel M, Bickel J E, Wolter B, Bergmann K V, Kubetzka A and Wiesendanger R 2013 \emph{Science} {\bf 341} 636

\bibitem{Yu_Skyrmion} Yu X Z, Kanazawa N, Onose Y, Kimoto K, Zhang W Z, Ishiwata S, Matsui Y and Tokura Y 2010 \emph{Nat. Mater.} {\bf 10} 106

\bibitem{Seki_Skyrmion} Seki S, Yu X Z, Ishiwata S and Tokura Y 2012 \emph{Science} {\bf 336} 198

\bibitem{Muller_Skyrmion} M$\ddot{u}$ller J, Rosch A and Garst M 2016 \emph{New J. Phys. } {\bf 18}  065006

	

\end{thebibliography}
\end{document}